\documentclass[10pt,journal]{IEEEtran}
\usepackage{times}
\usepackage[cmex10]{amsmath}
\usepackage{amssymb}
\usepackage{epsfig,verbatim}
\usepackage{algorithm}
\usepackage{algpseudocode}
\usepackage{cite}
\long\def\comment#1{}

\newfont{\bbb}{msbm10 scaled 700}

\newfont{\bb}{msbm10 scaled 1100}

\newcommand{\RR}{\mbox{\bb R}}

\newcommand{\ZZ}{\mbox{\bb Z}}
\newcommand{\FF}{\mbox{\bb F}}

\newcommand{\av}{{\bf a}}

\newcommand{\cv}{{\bf c}}

\newcommand{\ev}{{\bf e}}

\newcommand{\hv}{{\bf h}}

\newcommand{\qv}{{\bf q}}
\newcommand{\rv}{{\bf r}}

\newcommand{\uv}{{\bf u}}
\newcommand{\wv}{{\bf w}}

\newcommand{\xv}{{\bf x}}
\newcommand{\yv}{{\bf y}}
\newcommand{\zv}{{\bf z}}
\newcommand{\zerov}{{\bf 0}}

\newcommand{\Am}{{\bf A}}

\newcommand{\Hm}{{\bf H}}

\newcommand{\Qm}{{\bf Q}}

\newcommand{\Ac}{{\cal A}}

\newcommand{\Cc}{{\cal C}}

\newcommand{\Hc}{{\cal H}}
\newcommand{\Ic}{{\cal I}}

\newcommand{\Tc}{{\cal T}}
\newcommand{\Uc}{{\cal U}}

\newcommand{\Vc}{{\cal V}}

\newcommand{\zetav}{\hbox{\boldmath$\zeta$}}

\newcommand{\SNR}{{\sf SNR}}

\newcommand{\eqdef}{\stackrel{\Delta}{=}}

\newcommand{\transp}{{\sf T}}

\newtheorem{theorem}{Theorem}

\newtheorem{proposition}{Proposition}

\newcommand{\argmin}{\operatornamewithlimits{argmin}}

\title{MIMO Systems With Low-Resolution ADCs:\\Linear Coding Approach}

\author{
  \IEEEauthorblockN{Song-Nam Hong, Yo-Seb Jeon, and Namyoon Lee\\}
\thanks{ S.-N. Hong  is with the Department of Electrical Engineering, Ajou University, Suwon, Korea (e-mail: snhong@ajou.ac.kr). Y.-S. Jeon and N. Lee are with the Department of Electrical Engineering, POSTECH, Pohang, Gyeongbuk, Korea 37673 (e-mail: \{jys900311,nylee\}@postech.ac.kr). }
}

\begin{document}

\maketitle

%%%%%%%%%%%%%%%%%%
\begin{abstract}
This paper considers a multiple-input multiple-output (MIMO) system with low-resolution analog-to-digital converters (ADCs). In this system, the paper presents a new MIMO detection approach using coding theory. The principal idea of the proposed approach is to transform a non-linear MIMO channel to a linear MIMO channel by leveraging both a $p$-level quantizer and a lattice code where $p\geq 2$. After transforming to the linear MIMO channel with the sets of finite input and output elements, efficient MIMO detection methods are proposed to attain both diversity and multiplexing gains by using algebraic coding theory. In particular, using the proposed methods, the analytical characterizations of achievable rates are derived for different MIMO configurations. One major observation is that the proposed approach is particularly useful for a large MIMO system with the ADCs that use a few bits.
\end{abstract}

%%%%%%%%%%%%%%%%%%%%%%%%
\section{Introduction}
%%%%%%%%%%%%%%%%%%%%%%%%%%%%

There is an increasing demand for ultra-wideband communication systems to support hundreds of Gbps data rates for future wireless networks,
    because the system capacity can increase linearly with its bandwidth.
To implement communication systems that use a very large bandwidth, high speed analog-to-digital converters (ADCs) are indispensable.
As the speed of ADCs increases, however, it is very challenging to satisfy the power requirements of ADCs \cite{Murmann,Walden1999};
    specifically, the energy efficiency of ADCs dramatically drops when sampling rate is beyond 100MHz \cite{Walden1999}.
To reduce circuit complexity and power consumption, the use of very low-resolution ADCs (e.g., 1-5 bits) for ultra-wideband communication systems has received increasing attention over the past years \cite{Murmann,Walden1999,Nossek2006,Madhow2009,Madhow2010,Nossek2007,Nossek2008,Nossek2010}.

Once very-low-resolution ADCs are employed, the channel capacity is fundamentally limited by a quantization level.
In the extreme case when one-bit ADCs are used, QPSK modulation is information-theoretically optimal for the single-input single-output (SISO) fading channel \cite{Madhow2009},
    i.e., 2 bits/s/Hz is the maximum spectral efficiency in a communication system that uses the one-bit ADCs.
This limitation to spectral efficiency by the use of low-bit ADCs can be overcome using multiple antennas \cite{Larsson2014,Mo2015,Jacobsson2016,Wang2015,Choi2016};
for example, for a single-input multiple-output (SIMO)  channel, the spectral efficiency can increase logarithmically with the number of receive antennas,
    assuming the availability of perfect channel state information at the transmitter (CSIT) and perfect channel state information at receiver (CSIR) \cite{Mo2015}.

Recently, several authors have proposed specific symbol-detection algorithms for massive multiple-input multiple-output (MIMO) systems by incorporating the effect of low-resolution ADCs \cite{Choi2016,Liang2016,Mollen2016}.
For example, assuming perfect CSIR, a near-maximum-likelihood (nML) detection method was proposed for one-bit quantized signals \cite{Choi2016}.
Furthermore, the symbol-detection error of the MIMO systems with one-bit ADCs has been analyzed using linear-type detectors such as maximal ratio combining (MRC) or zero-forcing (ZF) when employing a least-squares channel estimator \cite{Larsson2014}.

Although MIMO detection algorithms that use one-bit ADCs are well understood due to their simplicity,
    efficient detection algorithms are not yet available for quantizers that use more than one bit.
The detection algorithms in \cite{Larsson2014,Mo2015,Jacobsson2016,Wang2015,Choi2016} have difficulty in increasing quantization bits,
    because the algorithms use {\em stair-type} (uniform or non-uniform) quantizers, and therefore create a non-linear MIMO channel with a finite set of output elements.
In such a non-linear MIMO channel, the computational complexity of the ML detector increases exponentially with both the number of transmit antennas and the quantization levels \cite{Choi2016}. The major limitation of the existing linear-type MIMO detection algorithms \cite{Larsson2014} is that they provide a reasonable performance only when the number of receive antennas is much larger than the number of transmit antennas.

%%%%%%%%%% Our Work %%%%%%%%%%
In this paper, we consider a MIMO system with $N_{\rm t}$ transmit antennas and $N_{\rm r}$ receive antennas.
Considering a $p$-level ADC per receive antenna ($p\geq 2$), we propose a different approach for developing efficient MIMO detection algorithms.
The key idea of the proposed approach is to transform a {\em non-linear} MIMO channel to a {\em linear} MIMO channel by using coding theory.
Specifically, instead of combating with a complicated non-linear MIMO channel induced by stair-type quantizer,
    we propose a modulo-type quantizer and a modulation/demodulation method based on lattice coding theory,
    which is motivated by the quantized compute-and-forward method introduced in \cite{Hong,LeeHong2016}.
By using finite numbers of outputs and inputs, this approach creates a linear MIMO channel over a finite field $\mathbb{Z}_p$.
After transforming to the linear MIMO channel with the sets of finite input and output elements, we propose efficient MIMO detection methods by using algebraic coding theory.

\begin{itemize}
\item We first consider the SIMO case, i.e., $N_{\rm t}=1$, to introduce the notion of receive diversity over finite-field operation when using limited ADCs.
    After the channel transformation, the SIMO system can be regarded as $N_{\rm r}$ parallel $p$-ary symmetric channels.
    Utilizing this fact, we present a simple receive-antenna-selection method that chooses the best subchannel with the minimum noise entropy (MNE).
    Then we provide a characterization of the achievable rate of the SIMO channel with $p$-level ADCs.
    The key idea of this analysis is to treat $N_{\rm r}$ quantized output signals as a codeword when repetition coding is used over a spatial domain.
    Using this result, then when $p=2$, we show that the proposed method is optimal under the assumption that the modulo-type quantizer is used in the SIMO system.
\item Next, we consider a symmetric MIMO channel, i.e., $N_{\rm t}=N_{\rm r}$.
    We present a successive-interference-cancellation (SIC) decoding algorithm when a family of {\em nested linear codes} is used for encoding.
    We demonstrate that the proposed method of encoding and decoding achieves the capacity of the transformed linear MIMO channel,
        provided that the quantized noise signals at the receive antennas are statistically independent.
    Furthermore, to reduce the decoding complexity, we introduce a simple ZF decoding method, which essentially inverses the linear channel matrix over the finite field.
    Although the proposed ZF method achieves a lower achievable rate than the previous one,
        the proposed method's computational complexity scales linearly with $p$, and is therefore particularly useful when $p$ is high.
\item Finally, we consider the asymmetric MIMO case in which $N_{\rm r}>N_{\rm t}$, and explain how to obtain both multiplexing and diversity gains in a finite-field MIMO system.
    With this purpose, we present a scheme that combines the proposed antenna selection method in Section IV,
        and then present the MIMO detection methods in Section V.
    The idea is to use the MNE criterion to exploit antenna selection diversity by selecting a set of $N_{\rm t}$ receive antennas among $N_{\rm r}$ antennas;
        this approach creates a symmetric MIMO channel.
    Then we apply the MIMO detection method introduced in Section V.
    Next, we generalize the scheme to obtain both diversity and multiplexing gains simultaneously.
    The key idea of the proposed scheme is to treat the finite-field MIMO channel as a generating matrix of a linear block code of block length $N_{\rm r}$ and code rate $\frac{N_{\rm t}}{N_{\rm r}}$.
    Using this scheme, we provide a characterization of an achievable rate of the MIMO system as a function of the minimum distance of the linear code when $N_{\rm r} > N_{\rm t}$.
\end{itemize}

\textbf{Notation:} Lower and upper boldface letters represent column vectors and matrices, respectively.
    For any two vectors $\xv$ and $\yv$ of the same length, $d_\mathsf{H}(\xv,\yv)$ represents Hamming distance which is the number of places at which they differ.
    Also, for any vector $\xv$, $w_\mathsf{H}(\xv)$ denotes Hamming weight, which is the number of nonzero locations in $\xv$.

%%%%%%%%%%%%%%%%%%%%%
\section{System Model}\label{sec:model}
\begin{figure}
\centerline{\includegraphics[width=9cm]{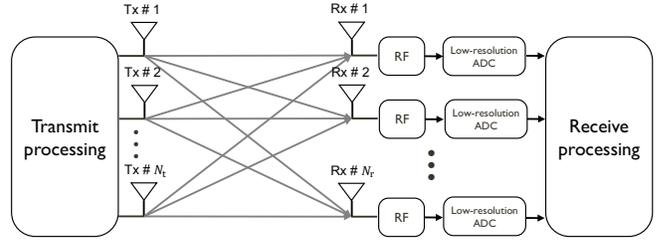}}
\caption{Illustration of a MIMO system with low resolution ADCs.}
\label{model}
\end{figure}

As illustrated in Fig. \ref{model}, we consider a MIMO system in which
a transmitter equipped with $N_{\rm t}$ antennas sends $N_{\rm t}$ information symbols to a receiver equipped with $N_{\rm r}$ antennas. Let ${\bf x} =[x_1, x_2,\ldots, x_{N_{\rm t}}]^{\top}$ be the channel input signal vector in which each element $x_i$ is uniformly selected from a constellation set $\Tc$. The received signal vector before ADC quantization, ${\bf y}=[y_1, y_2,\ldots, y_{N_{\rm r}}]^{\top}\in \mathbb{R}^{N_{\rm r}}$, is
\begin{align}
    {\bf y} = {\bf H}{\bf x} + {\bf z},
\end{align}
where ${\bf H}\in\mathbb{R}^{N_{\rm r}\times N_{\rm t}}$ denotes a channel matrix and ${\bf z}=[z_1, z_2,\ldots, z_{N_{\rm r}}]^{\top}\in \mathbb{R}^{N_{\rm r}}$ is real Gaussian noise with zero-mean and unit variance i.e., $\mathcal{N}(0,1)$. We assume that the channel remains constant during a finite resource block, i.e., a block fading model is assumed.
We also assume that ${\bf H}$ is known to the receiver.
In this paper, we only consider a real-valued channel for the ease of understanding of the proposed coding method,
    but it can be straightforwardly applied to a complex-valued channel by using the real-valued representation for complex vectors as
\begin{equation}
\left[ {\begin{array}{c}
   \mbox{Re}(\yv)  \\
   \mbox{Im}(\yv) \\
 \end{array} } \right]=\left[ {\begin{array}{cc}
   \mbox{Re}(\Hm) & -\mbox{Im}(\Hm) \\
   \mbox{Im}(\Hm) & \mbox{Re}(\Hm)\\
 \end{array} } \right]\left[ {\begin{array}{c}
   \mbox{Re}(\xv)  \\
   \mbox{Im}(\xv) \\
 \end{array} } \right]+\left[ {\begin{array}{c}
   \mbox{Re}(\zv)  \\
   \mbox{Im}(\zv) \\
 \end{array} } \right],
\end{equation}
where $\mbox{Re}(\av)$ and $\mbox{Im}(\av)$ denote the real and complex part of a complex vector $\av$, respectively.

%%%%%%%%%%%%%%%%%%%%%%%%%%%%%%%%%%%%

%%%%%%%%%%%%%%%%%%%%%%%
\section{The Proposed ADCs for a Lattice Code}\label{sec:main}

We present an ADC architecture and a modulation/demodulation method based on lattice theory, which is motivated by the quantized compute-and-forward introduced in \cite{Hong,LeeHong2016}.
This approach can transform a Gaussian MIMO channel with low-resolution ADCs into a linear MIMO channel over a finite-field,
whereas the use of a conventional stair-type ADC yields a non-linear MIMO channel.
Therefore, the use of the proposed ADC enables development of low-complexity MIMO detection methods when $N_{\rm t}$ and $N_{\rm r}$ are large; this result will be presented in the sequel.

%%%%%%%%%%%%%%%%
\subsection{Preliminaries}

For a lattice $\Lambda$, we define the lattice quantizer $Q_{\Lambda}(\xv)$ as the point of $\Lambda$ at minimum Euclidean distance from $\xv$,
    and the {\em Voronoi} region $\Vc$ of $\Lambda$ as the set of points $\xv$ such that $Q_\Lambda(\xv) = \zerov$ and
\begin{equation} \label{modulo}
    [\xv ]_{\Lambda} \eqdef [\xv] \mod \Lambda =  \xv - Q_{\Lambda}(\xv).
\end{equation}
Letting $\kappa \in \RR_+$, we consider the two nested one-dimensional lattices as
\begin{eqnarray}
    \Lambda_s  & = & \{x = \kappa p z: z \in \ZZ\} \nonumber \\
    \Lambda_c  & = & \{x = \kappa z: z \in \ZZ\}.
\end{eqnarray}
Here, $\kappa$ is chosen according to a transmit power constraint $\SNR$ as  $\kappa= \sqrt{2\SNR}$ for $p=2$ and $\kappa=\frac{\sqrt{12\SNR}}{p}$ for $p \geq 3$ as in \cite{LeeHong2016}.
Let $\ZZ_{p} = [\ZZ] \mod p\ZZ$ denote the finite-field of size $p$,
    with $p$ a prime number and $g : \ZZ_p \rightarrow \RR$ be a function that maps the elements of $\ZZ_p$ onto the points $\{0,...,p-1\} \subset \RR$.
Throughout the paper, $\oplus$  denotes an addition over a finite-field.

%%%%%%%%%%%%%%%%
\subsection{Lattice Modulation and Proposed ADC}

\begin{figure}
\centerline{\includegraphics[width=8.5cm]{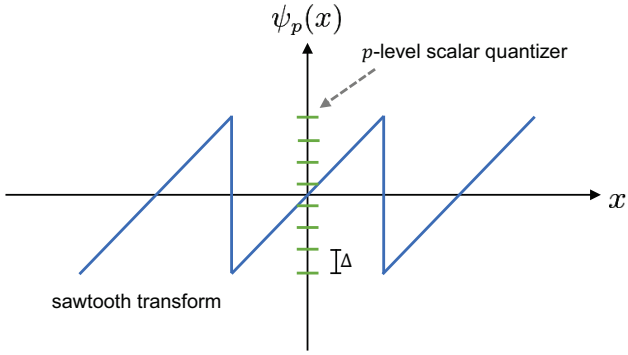}}
\caption{Illustration of the proposed ADC quantization.}
\label{ADC}
\end{figure}

Define the constellation set $\Tc \eqdef \Lambda_c \cap \Vc_s$, where $\Vc_s$ is the {\em Voronoi} region of $\Lambda_s$, i.e., the interval $[-\kappa p/2, \kappa p /2)$.
Then the lattice modulation mapping $\phi: \ZZ_p \rightarrow \Tc$ is defined as
\begin{align*}
    v &= \phi(u) \eqdef \left[\kappa g(u)\right]_{\Lambda_s}.
\end{align*}
The inverse function $\phi^{-1}(\cdot): \Tc \rightarrow \ZZ_p$ is referred to as the lattice demodulation mapping, and is given by
\begin{equation}
    u  = \phi^{-1}(v)\eqdef g^{-1}\left([v/\kappa] \mod p\ZZ \right),
\end{equation}
with $v \in \Tc$.
For an ADC, we propose a $p$-level scalar quantizer called a {\em sawtooth transform} as depicted in Fig.~\ref{ADC},
    which can be implemented by scalar quantization followed by a modulo operation as
\begin{equation}
    \psi_p(\cdot) = \left[Q_{\Lambda_c}(\cdot)\right]_{\Lambda_s}. \label{eq:ADC}
\end{equation}

%%%%%%%%%%%%%%%%%%%
\subsection{Building a Finite-Field MIMO Channel}

Each antenna $\ell$ transmits $x_{\ell}=\phi(c_\ell) \in \Tc$ with $c_\ell \in \ZZ_p$. Let $\hv_m^{\transp}$ denote the $m$-th row of $\Hm$.
Then the output signal of the $m$-th receiver antenna after ADC quantization is
\begin{align}
    \tilde{y}_m&= \psi_p\left(\hv_m^{\transp} \xv+z_m\right)=\left[Q_{\Lambda_c}\left( \hv_m^{\transp} \xv+z_m\right)\right]_{\Lambda_s}, \label{eq:ADC_output}
\end{align}
where the second equality is from the proposed ADC in (\ref{eq:ADC}) and $y_m \in \Tc$.

{\em Receiver operation:} The receiver first selects an {\em integer coefficient matrix} $\Am \in \ZZ^{N_{\rm r} \times N_{\rm t}}$ and then it produces the sequences of demodulated outputs as
\begin{equation}
    u_m = \phi^{-1}(\tilde{y}_m) \in \ZZ_p,\label{eq:demod_output}
\end{equation}
for $m \in \{1,...,N_{\rm r}\}$.
In detail, the output of ADC quantization is obtained as
\begin{align}
    \tilde{y}_m &=\psi_p\left(\sum_{\ell=1}^{N_{\rm t}} \Hm_{m,\ell}x_\ell +z_m\right)\nonumber\\
    &= \left[Q_{\Lambda_c}\left( \sum_{\ell=1}^{N_{\rm t}} \Hm_{m,\ell}x_\ell +z_m \right)\right]_{\Lambda_s} \nonumber\\
    &= \left[Q_{\Lambda_c}\left(\sum_{\ell=1}^{N_{\rm t}}\Am_{m,\ell}x_\ell + \sum_{\ell=1}^{N_{\rm t}} (\Hm_{m,\ell}-\Am_{m,\ell})x_\ell +z_m \right)\right]_{\Lambda_s} \nonumber\\
    &=\left[Q_{\Lambda_c}\left(\sum_{\ell=1}^{N_{\rm t}}\Am_{m,\ell}x_\ell + e_m +z_m \right)\right]_{\Lambda_s} \nonumber\\
    &= \left[\sum_{\ell=1}^{N_{\rm t}}\Am_{m,\ell}x_\ell\right]_{\Lambda_s} + \Big[Q_{\Lambda_c}\left( e_m +z_m \right)\Big]_{\Lambda_s},\label{eq:output}
\end{align}
where $e_m = \sum_{\ell=1}^{N_{\rm t}} (\Hm_{m,\ell}-\Am_{m,\ell})x_\ell $.
Applying lattice demodulation mapping to \eqref{eq:output} yields
\begin{align*}
    u_m &= \phi^{-1}(\tilde{y}_m) \\
    &= \phi^{-1}\left(\left[\sum_{\ell=1}^{N_{\rm t}}\Am_{m,\ell}\phi(c_\ell)\right]_{\Lambda_s} \right)\oplus  \phi^{-1}\Big(\left[Q_{\Lambda_c}\left( e_m +z_m \right)\right]_{\Lambda_s} \Big)\\
    &= \bigoplus_{\ell=1}^{N_{\rm t}} \Qm_{m,\ell}c_{\ell} \oplus \tilde{z}_{m},
\end{align*}
where a system matrix is defined as
\begin{align}
    \Qm_{m,\ell} &=g^{-1}([\Am_{m,\ell}] \mod p\ZZ), \label{eq:coef}
\end{align}
and an effective noise is defined as
\begin{align}
    \tilde{z}_m &= \phi^{-1}\left(\left[ Q_{\Lambda_c}(e_m + z_m)  \right]_{\Lambda_s} \right).\label{eq:e_noise}
\end{align}
Here, the marginal probability of mass function (pmf) of $\tilde{z}_m$ can be calculated as follows.
Letting $f_{m}(t)$ denotes the probability distribution of $e_m + z_m$, the pmf of $\tilde{z}_{m}$ is obtained as
\begin{equation}
P_{\tilde{z}_k}(u) = \int_{t \in \Vc_{\Phi(u)}} f_{k}(t)dt,
\end{equation}
where for $u \in \ZZ_p$ we define $\Vc_{\Phi(u)} = \{ y \in \RR: Q_{\Lambda_s}(y) = \Phi(u) + \kappa p m, \mbox{ for some } m \in \ZZ\}$.

{\em Finite-Field MIMO Channel:} Applying the above process  to all received signals, we finally obtain a linear MIMO channel over finite-field $\ZZ_p$ as
\begin{align}
\uv = \Qm\cv \oplus \tilde{\zv},\label{eq:fMIMO}
\end{align} where  $\cv=[c_1,\ldots,c_{N_{\rm t}}]^{\transp}$ denotes a transmit signal,
    $\uv=[u_1,\ldots,u_{N_{\rm r}}]^{\transp}$ denotes a receive signal,
    $\Qm$ represents the $N_{\rm r} \times N_{\rm t}$ channel matrix of which the $(m,\ell)$-th element $\Qm_{m,\ell}$ is defined in  \eqref{eq:coef},
    and the noise vector $\tilde{\zv}=[\tilde{z}_1,\ldots,\tilde{z}_{N_{\rm r}}]^{\transp}$ follows a joint distribution function $P_{\tilde{z}_1,\ldots,\tilde{z}_{N_{\rm r}}}$.
The capacity of the MIMO channel in \eqref{eq:fMIMO} is determined as a function of our choice of an integer coefficient matrix $\Am$.
Thus, the optimal choice of $\Am$ is to minimize the joint entropy of the noise vector (i.e., $\mathsf{H}(\tilde{z}_1,\ldots,\tilde{z}_{N_{\rm r}})$).
However, this approach does not lead to a tractable numerical method.
Instead, we resort to minimizing the variance of unquantized noise $e_m + z_m$ in (\ref{eq:e_noise}), as performed in \cite{Hong}.\\

%%%%%%%%%%%%%%%%%%%%%%%%%%%%%%%%%%%%

{\bf Remark 1:} When $p=2$, the proposed modulation method is to send a binary symbol from the set $\{0, \sqrt{2}\}$, i.e., on-off keying modulation.
Thus, the proposed method causes 3-dB shaping loss compared to the conventional binary phase shift keying (BPSK) modulation.
For $p\geq 3$, this shaping loss reduces to 1.24 dB compared to quadrature amplitude modulation (QAM) modulation.
As a result, while the proposed approach transforms the non-linear MIMO channel to the equivalent linear MIMO channel over a finite-field at the expense of the shaping loss in the modulation, and the loss is not significant when $p\geq 3$.

%{\bf Remark 2:} We have assumed that ${\bf H}$ is perfectly known to the receiver. To obtain the

%%%%%%%%%%%%%%%%%%%%%%%%%%%%%%%%%%%%%%%
%%%%%%%%%%%%%%%%%%%%%%%%%%%%%%%%%%%%%
\section{Finite-Field SIMO Channel: $N_{\rm t}=1$ and $N_{\rm r}\geq 2$}\label{sec:main2}

In this section we consider a special case of the MIMO channel in \eqref{eq:fMIMO} with $N_{\rm t} =1$.
In contrast to the conventional MIMO channel, it is not clear how to harness additional receiver observations in the finite-field SIMO channel.
The goal of this section is to introduce the notion of the receive diversity in the context of finite-field SIMO channels,
    and to illustrate how we can obtain two different types of receive diversity gains.
For a SIMO case, the input-output relationship in \eqref{eq:fMIMO} can be simplified to
\begin{equation}
    \uv = c \qv  \oplus \tilde{\zv}.
\end{equation}
This channel model can be seen as $N_{\rm r}$ parallel $p$-ary symmetric channels. Suppose that a single antenna of the receiver acts as a sub-channel for $\log{p}$-bit information transfer. This allows connection of $N_{\rm r}$ channels in parallel, where each sub-channel corresponds to one $p$-ary symmetric channel.
From this connection, the output signal of the $m$-th sub-channel is
\begin{equation}
    u_m = c q_m  \oplus \tilde{z}_m,\label{eq:N_r-parallel}
\end{equation}
assuming that $q_m \neq 0$ and that the values of $\tilde{z}_m$ are statistically independent.
If $\tilde{z}_m$ values are indeed statistically independent, the receiver is capable of exploiting $N_{\rm r}$ independent observations.

From \cite{Gamal}, the capacity of the above channel is given by
\begin{equation}
C = \mathsf{H}(u_1,\ldots,u_{N_{\rm r}}) - \sum_{i=1}^{N_{\rm r}} \mathsf{H}(\tilde{z}_{i}), \label{eq:capacity-SIMO}
\end{equation}
where $\mathsf{H}(\cdot)$ represents an entropy function.
This capacity can be achieved by a polar code \cite{Arikan2009,Sasoglu2009} as well as by a random linear code and joint decoding \cite{Gamal}.
Although a polar code is known to be a low-complexity coding scheme,
    it may not have low complexity when $N_{\rm r}$ becomes large (e.g., in massive MIMO systems),
    because the size of output alphabet grows exponentially with $N_{\rm r}$.
To reduce the complexity, the proposed approach operates in two steps:

{\em i) Output dimension reduction:} To reduce the complexity, define a function
\begin{equation}
    f(\cdot): p^{N_{\rm r}} \rightarrow p^{N_{\rm o}}
\end{equation}
for some integer $N_{\rm o} \geq 1$, where $N_{\rm o}$ does not grow with $N_{\rm r}$.
In this paper, we restrict ourselves to choose $N_{\rm o} = N_{\rm t}$ and leave the case of $N_{\rm o} > N_{\rm t}$ for future work.
In the SIMO case, this function maps an $N_{\rm r}$-dimensional output vector to a $p$-ary signal, i.e.,
\begin{equation}
\hat{\uv} = f(\uv) \in \ZZ_{p}.
\end{equation}
This function can be interpreted as a combiner to obtain receive diversity in the finite-field SIMO system.
The optimal receiver combining function can be obtained by solving
\begin{align}
    &\mbox{maximize } \mathsf{I}(c;f(\uv)),
\end{align} where $\mathsf{I}(A ; B)$ represents the mutual information between two random variables $A$ and $B$. Unfortunately, this  optimization problem is sophisticated to obtain a closed-form solution. In the sequel, we will present two methods to design a receive combining function $f(\cdot)$ in Sections~\ref{subsec:AS} and~\ref{subsec:diversity}.

{\em ii) Channel coding:} After applying the receive combining function to observation $\uv$,
    we can yield a point-to-point channel with input $c \in \ZZ_p$ and output $\hat{u}=f(\uv) \in \ZZ_p$.
Using a capacity-achieving outer code (e.g., polar code), we can achieve the capacity of the resulting point-to-point channel,
    where the complexity of polar code is independent of $N_{\rm r}$.
Then the achievable rate of the proposed scheme is
\begin{equation}
    R=\mathsf{I}(c;f(\uv)).
\end{equation}

%%%%%%%%%%%%%%%%%%%%%%%%%%%%%%%%%%%%
\subsection{Antenna Selection with Minimum Noise Entropy (MNE)} \label{subsec:AS}

One intuitive strategy to obtain receive diversity gains is {\em antenna selection} for the finite-field SIMO system.
To get an intuitive understanding of the proposed antenna selection method, we recall a conventional antenna-selection strategy for a Gaussian SIMO channel.
The key principle of antenna selection is to choose the receive antenna with the highest channel gain between it and a transmitter antenna to receiver antennas.
Intuitively, this simple strategy increases the achievable rate due to channel diversity, provided that the channel gains are not perfectly correlated.
Accordingly, in the finite-field SIMO system, the receiver is capable of observing $N_{\rm r}$ different output signals, each of which experiences a different noise entropy.
Our antenna-selection strategy is to choose the sub-channel that yields the highest sub-channel capacity.
From the $N_{\rm r}$ parallel $p$-ary symmetric channels in \eqref{eq:N_r-parallel}, it is well known that the capacity of the $m$-th channel is
\begin{equation}
C_m = \log{p} - \mathsf{H}(\tilde{z}_m).
\end{equation}
Because selecting the best sub-channel with the highest channel capacity is equivalent to choosing the sub-channel with the minimum noise entropy,
    the proposed antenna-selection strategy is to identify a receive antenna index $m^\dag$ such that
\begin{equation}
m^\dag = \argmin_{m \in \{1,...,N_{\rm r}\}} \mathsf{H}(\tilde{z}_m).\label{eq:opt_m}
\end{equation}
For the proposed antenna-selection method, we can define the output combiner function as a linear vector $\ev_{m^{\dag}}$ in which the $m^\dag$th component is 1, and all other elements are $0$, i.e.,
\begin{equation}
f(\uv) = \ev_{m^\dag}^{\transp}\uv.
\end{equation}
As a result, the achievable rate of the proposed antenna selection strategy for the finite-field SIMO channel is
\begin{equation}
R_{\mbox{{\tiny AnSe}}} = \log{p} - \mathsf{H}(\tilde{z}_{m^\dag}),
\end{equation}
where $m^\dag$ is the solution of the optimization problem \eqref{eq:opt_m}.
The following lemma shows that the proposed antenna selection strategy is optimal when restricted to a linear function $f(\cdot)$ where
\begin{equation}
    f(\uv) = \wv^{\transp}\uv
\end{equation} for some vector $\wv \in \ZZ_{p}^{N_{\rm r}}$.

%%%%%%%%%%%%%%%%%%%%%%%%%%%%%%%%%%%%%%%%%%%%%%%%%%%%%%%%%%%
\begin{proposition}\label{prop:antenna}
    The proposed antenna selection strategy is the optimal linear scheme.
\end{proposition}
\begin{IEEEproof}
    Let $m^\dag$ denote the solution of the proposed antenna selection strategy from \eqref{eq:opt_m}.
    Then, for any combining vector $\wv$,
    \begin{align*}
        \mathsf{I}\left(c~;~\bigoplus_{i=1}^{N_{\rm r}}w_i u_i\right) &= \log{p} - \mathsf{H}(w_1\tilde{z}_1 \oplus \cdots \oplus w_{N_{\rm r}}\tilde{z}_{N_{\rm r}})\\
        &\stackrel{(a)}{\leq}  \log{p} - \min\{\mathsf{H}(\tilde{z}_i): w_{i}=1\}\\
        &\stackrel{(b)}{=} \log{p} - \mathsf{H}(\tilde{z}_{m^\dag}),
    \end{align*}
    where (a) is due to the fact that $\mathsf{H}(\tilde{z}_i)$ is obtained by adding conditions to $\mathsf{H}(a_1\tilde{z}_1 \oplus \cdots \oplus a_{N_{\rm r}}\tilde{z}_{N_{\rm r}})$ and  the condition decreases the entropy, and (b) is by the definition of the proposed antenna-selection strategy. This completes the proof.
\end{IEEEproof}
%%%%%%%%%%%%%%%%%%%%%%%%%%%%%%%%%%%%%%%%%%%%%%%%%%%%%%%%%%%%%%%%%%%%%%

\begin{figure}
\centerline{\includegraphics[width=9cm]{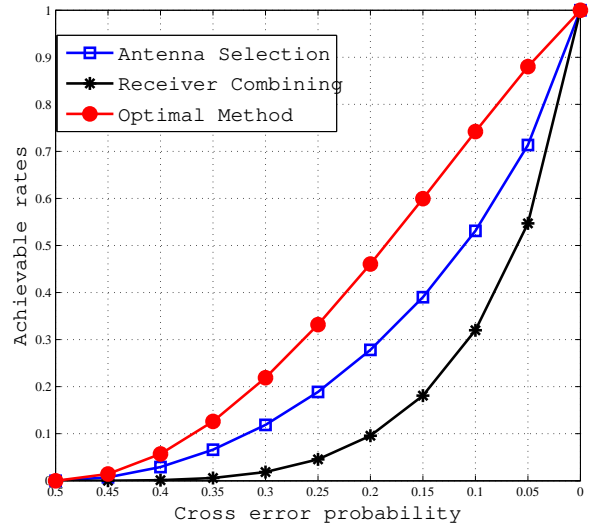}}
\caption{$N_{\rm r}=2$. Achievable rates of various receive diversity schemes for a SIMO system over $\ZZ_2$ where both sub-channels have the same cross-probability. }
\label{toy-exam}
\end{figure}

From Proposition~\ref{prop:antenna}, we observe that when restricted to a linear function $f$ increasing the number of observations can degrade the performance in the finite-field channel,
    which is totally different from conventional Gaussian channels.
As can be seen in Fig.~\ref{toy-exam}, the proposed antenna-selection scheme provides a better achievable rate at all cross error probabilities that does the linear combining method in (25). Although the proposed antenna-selection scheme is the optimal linear scheme, its achievable rate is still far below optimal, which is obtained by the numerical computation of (16).
Thus, a low-complexity non-linear function $f$ should be considered; we do this in Section~\ref{subsec:diversity}.

%%%%%%%%%%%%%%%%%%%%%%%%%%%%%%%%%%%%%%%%%%%%%%%%%%
%%%%%%%%%%%%%%%%%%%%%%%%%%%%%%%%%%%%%%%%%%%%%%%%%%%

%%%%%%%%%%%%%%%%%%%%%%%%%%%%%%%%%%%%%%%%%%%%%%%%%
\subsection{Repetition Coding over Receive Antennas}\label{subsec:diversity}

We present a new receive diversity technique in a finite-field SIMO system, by using a simple repetition coding over a spatial domain.
This coding can be represented as a non-linear function $f(\cdot)$.
To explain the main idea of the proposed coding method, we first assume a binary case ($p=2$).

Consider the simple case of $N_{\rm r} = 3$.
Suppose that transmitter sends $c=1$ and the received output vector is  $\uv = [u_1,u_2,u_3]^{\transp}=[1,1,0]^{\transp}$.
The condition $u_m = c \oplus \tilde{z}_{m}$ for $m \in \{1,2,3\}$ implies that the first two sub-channels are good while noise in the last sub-channel flips the output (i.e., $\tilde{z}_1 = \tilde{z}_2 = 0$ and $\tilde{z}_3 = 1$).
Because the same information was sent through the three different sub-channels,
    it is natural to determine the output of the combiner function $\hat{u} = f(\uv) = 1$ by the majority decoding principle \cite{MacWilliams}.
By constructing the combiner function $f(\cdot)$ based on the majority decoding rule, the receiver is capable of correcting at least one error among the three sub-channel outputs.
Let $\hat{u} = f(\uv)$ be the output obtained by the majority decoding function.
Then the achievable rate is
\begin{equation}
R_{\mbox{{\tiny Rep}}} = 1 - \mathsf{H}_2(P_\epsilon),
\end{equation} where $\mathsf{H}_2(\alpha) = - \alpha\log\alpha - (1-\alpha)\log(1-\alpha)$ denotes a binary entropy function and $P_{\epsilon}$ denotes the error-probability under majority decoding (which is equivalent to the cross-probability of the corresponding BSC with input $c$ and output $\hat{u}$) as
\begin{equation}
P_{e} = (1-\epsilon_1)\epsilon_2\epsilon_3 + (1-\epsilon_2)\epsilon_1\epsilon_3 + (1-\epsilon_3)\epsilon_1\epsilon_2 + \epsilon_1\epsilon_2\epsilon_3,
\end{equation} where $\epsilon_m = \mathbb{P}(\tilde{z}_m = 1)$ for $m \in \{1,2,3\}$.

Now, we consider a general case with $N_{\rm r}$ receive antennas.
The majority decoding function can be defined as
\begin{equation}
\hat{u} = f(\uv) = \left\lfloor \frac{1}{2}+\frac{\sum_{m=1}^{N_{\rm r}}u_m -\frac{1}{2}}{N_{\rm r}} \right\rfloor.
\end{equation}
This decoding function is able to correct at least $\left\lfloor \frac{N_{\rm r} - 1}{2} \right \rfloor$ errors among $N_{\rm r}$ observations $\uv$.
We define a $k$-combination of set $\{1,...,N_{\rm r}\}$ as a subset of $k$ distinct elements of $\{1,...,N_{\rm r}\}$.
Therefore, ${{N_{\rm r}}\choose{k}}$ $k$-combination subsets exist.
Let $\mathcal{S}_{k,j} =\{\pi_j(1),\pi_j(2),\ldots, \pi_j(k)\}$ with $|\mathcal{S}_{k,j}|=k \leq N_{\rm r}$ be the $j$th $k$-combination subset.
Here, $\pi_j(m)$ is the $m$th element of the $j$th $k$-combination subset.
Using this notation, the effective error-probability of the proposed receive diversity scheme with asymmetric error probabilities per sub-channel is
\begin{align}
P_{e}(N_{\rm r}) = \sum_{k=\lfloor\frac{ N_{\rm r}-1}{2}\rfloor+1}^{N_{\rm r}}  \sum_{j=1}^{{{N_{\rm r}}\choose{k}} } \prod_{\ell \in \mathcal{S}_{k,j}}\epsilon_{\ell}  \prod_{m \in \mathcal{S}^c_{k,j}} (1-\epsilon_m),
\end{align}
where $ \mathcal{S}^c_{k,j} =\{1,...,N_{\rm r}\}\setminus\mathcal{S}_{k,j}$. Therefore the achievable rate is
\begin{align}
R_{\mbox{{\tiny Rep}}}&=1-\mathsf{H}_2(P_{e}(N_{\rm r})).
\end{align}
Because this expression is complicated, we present an example that clearly shows receive diversity gains as a function of $N_{\rm r}$.

Under the premise that the noise entropy is identical in all sub-channels, i.e., $\epsilon_m = \epsilon$ for all $m \in \{1,...,N_{\rm r}\}$,
    the effective cross-probability of the BSC with input $c$ and output $\hat{u}$ is
\begin{align}
P_{e,{\rm sym}}(N_{\rm r}) = \sum_{j=\lfloor\frac{ N_{\rm r}-1}{2}\rfloor+1}^{N_{\rm r}}  {{N_{\rm r}}\choose{j}} \epsilon^j(1-\epsilon)^{N_{\rm r}-j}.
\end{align}
Consequently, the achievable rate of this symmetric case is
\begin{align}
R_{\mbox{{\tiny Rep, sym}}}&=1-\mathsf{H}_2(P_{e,{\rm sym}}(N_{\rm r})).
\end{align}
This evidently demonstrates that the proposed majority-decoding method over the spatial domain increases the achievable rate as $N_{\rm r}$ increases,
    because $P_{e,{\rm sym}}(N_{\rm r})$ is a decreasing function of $N_{\rm r}$.
This trend can be understood as receive diversity gain in a finite-field SIMO channel.\\

\begin{figure}
\centerline{\includegraphics[width=9cm]{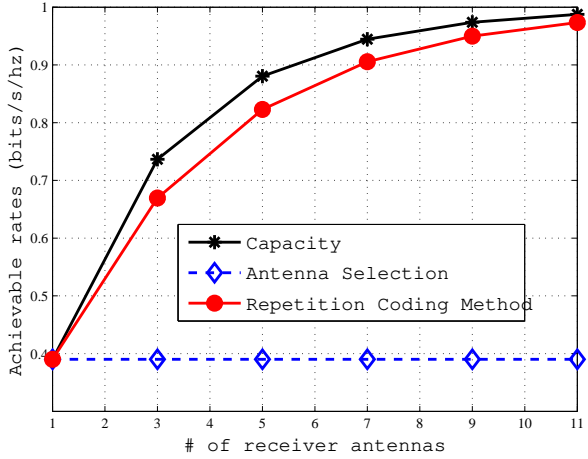}}
\caption{Achievable rates of the proposed antenna selection and repetition coding method for $p=2$ and $\epsilon=0.15$.}
\label{capacity-comp}
\end{figure}

%%%%%%%%%%%%%%%%%%%%%
{\bf Example 1:} We compare the achievable rates of the proposed antenna selection strategy and the repetition coding method as a function of $N_{\rm r}$.
For simplicity, we assume that $p=2$, $\mathbb{P}(\tilde{z}_m=1) = \epsilon$ for $m \in \{1,...,N_{\rm r}\}$, and that $N_{\rm r}$ is an odd number.
Then the capacity of this channel in (\ref{eq:capacity-SIMO}) can be simply computed as
\begin{align}
    C=\sum_{j=0}^{ N_{\rm r}  }  \left(\begin{array}{c}  N_{\rm r} \\ j \end{array}\right) (-\alpha_j \log{\alpha_j}) - N_{\rm r} \mathsf{H}_2(\epsilon),
\end{align}
where $\alpha_j =\frac{1}{2}\left(\epsilon^{N_{\rm r} - j}(1-\epsilon)^{j} + \epsilon^{j} (1-\epsilon)^{N_{\rm r} -j }\right)$.
This formula is used to plot the capacities in Fig.~\ref{capacity-comp}.
In Fig.~\ref{capacity-comp}, the repetition coding method achieves higher achievable rate than the antenna selection strategy, and almost achieves theoretical capacity.
Further, both the capacity and the achievable rate of the repetition coding method improve logarithmically as $N_{\rm r}$ increases.\\ %\end{example}

%%%%%%%%%%%%%%%%%%%%%%%%%%%%%%%%%%%%%%%%%%%%%%%%%%%%%%%%
%%%%%%%%%%%%%%%%%%%%%%%%%%%%%%%%%%%%%%%%%%%%%%%%%%%%%%%%%

\begin{figure}
\centerline{\includegraphics[width=9cm]{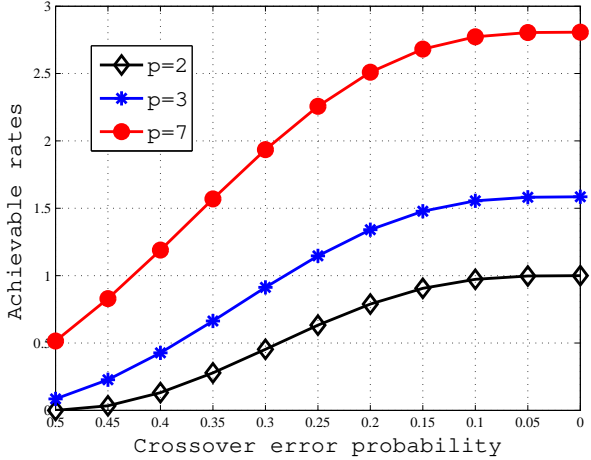}}
\caption{$N_{\rm r}=7$. Achievable rates of the proposed repetition coding method as a function of field-size $p$.}
\label{capacity-non}
\end{figure}

Generalizing the above coding method into an arbitrary prime $p$, we obtain the following theorem:
%%%%%%%%%%%%%%%%%%%%%%%%%%%%%%%%%%%%%%%%%%%%%%%
\begin{theorem}\label{thm:SIMO}
    Consider a SIMO channel with a channel coefficient vector $\qv=[q_1,\ldots,q_{N_{\rm r}}]^{\transp} \in \ZZ_p^{N_{\rm r}}$ with $q_m \neq 0$ for $m \in \{1,...,N_{\rm r}\}$ where  $\mathbb{P}(\tilde{z}_m \neq 0) = \epsilon_m$, for  $m \in \{1,...,N_{\rm r}\}$.
    Then the repetition coding method achieves the rate of
    \begin{equation*}
    R_{\mbox{{\tiny Rep}}}=\log{p} - \mathsf{H}_2(P_{e}(N_{\rm r})) - P_{e}(N_{\rm r})\log{(p-1)},
    \end{equation*}where
    \begin{equation*}
    P_{e}(N_{\rm r}) = \sum_{k=\lfloor\frac{ N_{\rm r}-1}{2}\rfloor+1}^{N_{\rm r}}  \sum_{j=1}^{{{N_{\rm r}}\choose{k}} } \prod_{\ell \in \mathcal{S}_{k,j}}\epsilon_{\ell}  \prod_{m \in \mathcal{S}^c_{k,j}} (1-\epsilon_m).
    \end{equation*}
\end{theorem}
%%%%%%%%%%%%%%%%%
\begin{IEEEproof}
    The decoding procedures exactly follow the binary case in the above.
    Then the receiver performs the majority decoding with an observation
    $\uv=[u_1,\ldots,u_{N_{\rm r}}]^{\transp}$, i.e.,
    \begin{equation}
    \hat{u} = f(\uv) \in \ZZ_p.
    \end{equation}
    Based on this, we define the transition probabilities as
    \begin{align}
    \beta_{i,x} = \mathbb{P}(\hat{u} = x \oplus i | c=x).
    \end{align}
    Due to the symmetry of the channel, it can be easily shown that $\beta_{i,0} = \beta_{i,x}$ for any $x \in \ZZ_{p}$,
        so for ease of notation, we drop the subscript $x$ as $\beta_{i}=\beta_{i,x}$ for all $x \in \ZZ_p$.
    Similar to the binary case, we can yield a point-to-point channel:
    \begin{align}
    \hat{u} = c \oplus \zeta
    \end{align}
    where $\mathbb{P}(\zeta = i ) = \beta_i$ for $i \in \ZZ_p$.
    Then this scheme achieves the rate of
    \begin{align}
    R=\log{p} - \mathsf{H}(\zeta). \label{eq:cap}
    \end{align}
    Here, the computation of $\mathsf{H}(\zeta)$ is quite complicated especially for a large $N_{\rm r}$.
    Instead, we will compute the lower bound of $\mathsf{H}(\zeta)$.
    We first compute the error probability as
    \begin{align}
    1-\beta_0 &\leq \mathbb{P}(w_\mathsf{H}(\tilde{\zv})> t) \nonumber \\
    &= \sum_{k=t+1}^{N_{\rm r}}  \sum_{j=1}^{{{N_{\rm r}}\choose{k}} } \prod_{\ell \in \mathcal{S}_{k,j}}\epsilon_{\ell}  \prod_{m \in \mathcal{S}^c_{k,j}} (1-\epsilon_m),\label{eq:error-upper}
    \end{align}
    where $t = \lfloor\frac{ N_{\rm r}-1}{2}\rfloor$.
    Contrast to the binary case, equation  (\ref{eq:error-upper}) is an upper bound on the error probability because, for some error patterns $\ev$ with $w_\mathsf{H}(\ev)> t$,
        the majority decoding can provide a valid output.
    For example, when $p=5$ and $N_{\rm r} = 5$, the received signal $[0,0,1,2,3]^{\transp}$ can be correctly decoded as $0$ although the number of errors is 3 (larger than $t=2$).
    Note that (\ref{eq:error-upper}) is equal to $P_{e}(N_{\rm r})$.  We let $\beta_0' =1- P_{e}(N_{\rm r})$ and $\beta_{i}' = \frac{1-P_{e}(N_{\rm r})}{p-1}$ for $i \in \{1,...,p-1\}$.
    We define a random variable $\zeta'$ with $\mathbb{P}(\zeta' = i) = \beta_i'$ for $i\in \{0,...,p-1\}$,
    then we can obtain the lower bound of (\ref{eq:cap}) as
    \begin{align}
    R&=\log{p} - \mathsf{H}(\zeta) \nonumber \\
    &\stackrel{(a)}{\geq} \log{p} - \mathsf{H}(\zeta') \nonumber\\
    &=\log{p} - \mathsf{H}_2(P_{e}(N_{\rm r})) - P_{e}(N_{\rm r})\log{(p-1)}, \label{eq:non-binary-bound}
    \end{align}
    where (a) is due to the fact that $\beta_0 \geq \beta_0'$ and uniformization of a probability distribution can increase the entropy. This completes the proof.
\end{IEEEproof}
\vspace{0.2cm}
%%%%%%%%%%%%%%%%%%%%%%%%%
{\bf Example 2:} Fig.~\ref{capacity-non} shows the achievable rates of the proposed repetition coding method for various field-size $p$.
In Fig.~\ref{capacity-non}, as the error probability decreases, the achievable rate is increased by increasing $p$.
This is equivalent to increasing the order of modulation in communication systems as $\SNR$ increases.
Thus, we should choose an appropriate field size $p$ by considering the tradeoff between achievable rate and complexity.\\

%%%%%%%%%%%%%%%%%%%%%%%%%

%%%%%%%%%%%%%%%%%%%%%%%%%%%%%%%%%%%%%%%%%%%
%%%%%%%%%%%%%%%%%%%%%%%%%%%%%%%%%%%%%%%%%%

%%%%%%%%%%%%%%%%%%%%%%%%%%%%%%%%%%%%
\section{Finite-Field MIMO Channel: $N_{\rm t} = N_{\rm r}$}\label{sec:main1}

We consider a MIMO channel over $\ZZ_p$, defined in (\ref{eq:fMIMO}) with $N_{\rm t}=N_{\rm r}$.
For such a channel, the sum-capacity is obtained \cite[Theorem 4]{Hong} as
\begin{equation}
C_{{\rm sum}} = N_{\rm r}\log{p} - \mathsf{H}(\tilde{z}_{1},\ldots,\tilde{z}_{N_{\rm r}}),\label{eq:capacity}
\end{equation}
where $\mathsf{H}(\cdot)$ denotes the joint entropy of random variables.
This capacity can be achieved by a random linear codebook and a joint decoding  \cite{Gamal} but its complexity is not manageable.
Thus, in the next subsections, we present two low-complexity schemes: Matrix inversion (a.k.a., zero-forcing receiver) and successive coding,
    then show that the successive coding can achieve the capacity in (\ref{eq:capacity}) if $\tilde{z}_m$ values are statistically independent.

%%%%%%%%%%%%%%%%
\subsection{Successive Coding}\label{subsec:SC}

The idea of successive coding (SC) is to combine a MIMO receiver with channel coding efficiently.
For the ease of explanation, we assume that
\begin{equation}
\mathsf{H}(\tilde{z}_{1}) \leq \mathsf{H}(\tilde{z}_{2}) \leq \cdots \leq \mathsf{H}(\tilde{z}_{N_{\rm r}}).
\end{equation}
We use a family of {\em nested linear codes} $\Cc_{1} \supseteq \Cc_{2} \supseteq \cdots \supseteq \Cc_{N_{\rm r}}$ of respective code rates $r_{1}\geq r_{2} \geq \cdots \geq r_{N_{\rm r}}$. Each member of the family can achieve the capacity of the underlying symmetric channel \cite{Dobrushin}.
To transmit the $m$th stream, we employ the linear code $\Cc_m$. Let $\cv_{m} \in \Cc_{m}$ denote the codeword corresponding to the $m$th stream.
Due to the use of nested linear codes,
\begin{equation}
\cv_{m} \in \Cc_{k} \mbox{ for all } k \leq m.
\end{equation}

{\em Decoding procedures:} From the $u_1= \bigoplus_{\ell=1}^{N_{\rm r}} \Qm_{1, \ell} \cv_{\ell} \oplus \tilde{z}_1$, the receiver first decodes the linear combination of codewords as
\begin{equation}
    \cv_{1}' = \bigoplus_{\ell=1}^{N_{\rm r}} \Qm_{1, \ell} \cv_{\ell},
\end{equation}
where $\cv_{1}' \in \Cc_1$.
Then the receiver can successfully decode the $\cv_{1}'$ if
\begin{equation}
    r_{1} \leq \log{p} - \mathsf{H}(\tilde{z}_1).
\end{equation}
Using the $\cv_1'$ previously decoded, the receiver can eliminate the term $\cv_1$ from $u_2 = \bigoplus_{\ell=1}^{N_{\rm r}} \Qm_{2,\ell}\cv_{\ell} \oplus \tilde{z}_2$ as
\begin{equation}
    u_{2}' = u_2 \ominus \Qm_{1,1}^{-1}\Qm_{2,1} \cv_1 = \cv_{2}' \oplus \tilde{z}_{2},
\end{equation}
where $\ominus$ denotes subtraction over $\ZZ_p$ and we let $\cv_{2}' = \bigoplus_{\ell=2}^{N_{\rm r}} \Qm_{2,\ell}'\cv_{\ell}$.
Because $\cv_{2}'  \in \Cc_{2}$, the receiver can successfully decode it if
\begin{equation}
    r_{2} \leq \log{p} - \mathsf{H}(\tilde{z}_2).
\end{equation}
By repeatedly applying the above procedures, the receiver can decode the $N_{\rm r}$ codewords  $\cv_{m}'=\bigoplus_{\ell=m}^{N_{\rm r}} \Qm_{m,\ell}'\cv_{\ell}  \in \Cc_{m}$ for $m \in [N_{\rm r}]$, if the following $N_{\rm r}$ constraints are satisfied:
\begin{equation}
    r_{m} \leq \log{p} - \mathsf{H}(\tilde{z}_{m}), \mbox{ for all }m \in \{1,...,N_{\rm r}\}.
\end{equation}
From the $N_{\rm r}$ decoded codewords $\cv_{m}'$ for $m \in \{1,...,N_{\rm r}\}$, the receiver can obtain the desired codewords $\cv_m$ for $m \in \{1,...,N_{\rm r}\}$ by using simple matrix inversion with $\Qm'$.
This operation is possible because $\Qm'$ has a full rank as long an $\Qm$ has a full rank.
Then this scheme can achieve the sum-rate of
\begin{equation}
R_{\mbox{{\tiny SC}}} = N_{\rm r} \log{p} - \sum_{m=1}^{N_{\rm r}}\mathsf{H}(\tilde{z}_m).\label{eq:ach-SC}
\end{equation}
Based on this observation, we have:

%%%%%%%%%%%%%%%%%%%%%%%%
\begin{proposition}
    The proposed SC achieves the capacity of the MIMO channel in (\ref{eq:fMIMO}) if $\tilde{z}_m$ values are statistically independent.
\end{proposition}
\begin{IEEEproof}
    Because $\tilde{z}_m$ values are assumed to be statistically independent,
    the capacity in (\ref{eq:capacity}) is equal to
    \begin{equation}
    C= N_{\rm r}\log{p} - \sum_{m=1}^{N_{\rm r}} \mathsf{H}(\tilde{z}_{m}).
    \end{equation}
    From (\ref{eq:ach-SC}), we can see that this capacity is achieved by the SC. This concludes the proof.
\end{IEEEproof}

%%%%%%%%%%%%%%%
\subsection{Zero-Forcing }\label{subsec:MI}

For Gaussian channels, linear MIMO receivers such as ZF and minimum mean square error (MMSE) have been widely used due to their satisfactory sum rates and low complexities.
Therefore, in this section, we consider the ZF method for a finite-field MIMO channel in which the goal of ZF is to eliminate the all interferences as
\begin{equation}
    \hat{\uv} = \Qm^{-1}\uv = \cv \oplus \Qm^{-1}\tilde{\zv}.
\end{equation}
This yields $N_{\rm r}$ parallel (interference-free) channels as
\begin{equation}
    \hat{u}_m= c_m+ \zeta_m
\end{equation}
for $m \in \{1,...,N_{\rm r}\}$, where
\begin{equation}
    \zeta_m = \bigoplus_{i=1}^{N_{\rm r}} \Qm^{-1}_{m,i}\tilde{z}_{i}.
\end{equation}
Applying a capacity-achieving linear code (e.g., polar code) to each channel independently, this scheme achieves the sum-rate of
\begin{equation}
    R_{\mbox{{\tiny ZF}}} = N_{\rm r}\log{p} - \sum_{m=1}^{N_{\rm r}} \mathsf{H}(\zeta_{m}).\label{eq:ach-MI}
\end{equation}
This observation implies

\begin{proposition}
    The SC achieves higher sum rate than ZF, i.e.,
    \begin{equation*}
        R_{\mbox{{\tiny SC}}} \geq R_{\mbox{{\tiny ZF}}}.
    \end{equation*}
\end{proposition}
\begin{IEEEproof}
    Because the condition decreases the entropy, we have
    \begin{align*}
    \mathsf{H}(\zeta_{m}) &\geq \mathsf{H}(\zeta_{m}|\{\tilde{z}_{i}: i \in \{1,\ldots,N_{\rm r}\}\setminus \{m\}\})\\
    & = \mathsf{H}(\tilde{z}_{m}),
    \end{align*}
    which shows that $ \sum_{m=1}^{N_{\rm r}} \mathsf{H}(\zeta_{m}) \geq  \sum_{m=1}^{N_{\rm r}}\mathsf{H}(\tilde{z}_m)$. This completes the proof.
\end{IEEEproof}

%%%%%%%%%%%%%%%%%%%%%%%%%%%%%%%%
\section{Finite-Field MIMO Channel: $N_{\rm r}>N_{\rm t}$}

In this section, we present a method to attain both multiplexing and diversity gains in a finite-field MIMO system.
Unlike the SIMO case, it is unclear how to obtain diversity gains in the MIMO case,
    because multiple data streams are simultaneously transmitted by a transmitter and they interfere with each other at the receiver.

%%%%%%%%%%%%%%%%%%%%%%%%%%%%%%%%
\subsection{Proposed Scheme That Uses Antenna Selection}\label{subsec:AS}

This scheme consists of antenna selection and successive decoding as
\begin{itemize}
    \item Find $N_{\rm t}$ observations among $N_{\rm r}$ observations such that the sum-capacity of the resulting $N_{\rm t} \times N_{\rm t}$ MIMO channel is maximized.
    \item Perform the successive coding in Section~\ref{subsec:SC} on the resulting MIMO channel.
\end{itemize}
In this section, we will explain how to choose $N_{\rm t}$ observations (i.e., antenna-selection strategy),
    because the second part exactly follows the procedures in Section~\ref{subsec:SC}.
The antenna selection problem can be represented as
\begin{align}
    \Uc^{*}&=\argmin_{\Uc \subset \{1,...,N_{\rm r}\}}   \sum_{i \in \Uc} \mathsf{H}(\tilde{z}_i) \label{eq:opt}\\
    &\mbox{subject to } \mbox{Rank}\left(\Qm(\Uc,\{1,...,N_{\rm t}\})\right) = N_{\rm t}. \nonumber
\end{align}
This problem consists of the minimization of linear function subject to a matroid constraint,
    where the matroid $(\Omega,\Ic)$ is defined by the ground set $\Omega = \{1,...,N_{\rm r}\}$ and by the collection of independent sets $\Ic = \{\Uc\subseteq \Omega: \Qm(\Uc,\{1,...,N_{\rm t}\}) \mbox{ has linearly independent rows}\}$.
A greedy algorithm finds an optimal solution, where at each step, the index that corresponds to the minimum $\mathsf{H}({\tilde{z}_i})$ among the values
    of which the corresponding row is linearly independent of the rows previously chosen  \cite{Rado,Edmonds}.
Applying the successive coding in Section~\ref{subsec:SC} to the selected observations yields the sum-rate of
\begin{equation}
    R_{\mbox{\tiny AnSe}} = N_{\rm t} \log{p} - \sum_{m \in \Uc^*} \mathsf{H}(\tilde{z}_m),\label{eq:ach_AS}
\end{equation}
where $\Uc^*$ denotes the optimal solution of the optimization problem in (\ref{eq:opt}).

%\begin{figure}
%\centerline{\includegraphics[width=9cm]{non-binary-dmin}}
%\caption{Performance of the proposed coding approach for a MIMO system as a function of minimum distance.}
%\label{capacity-comp}
%\end{figure}

%%%%%%%%%%%%%%%%%%%%%%%%%%%%%%%%%%
\subsection{Proposed Scheme That Uses Linear Block Codes}\label{subsec:LBC}

The key idea of the proposed scheme is to treat the MIMO channel matrix $\Qm=[\qv_1,\qv_2,\ldots,\qv_{N_{\rm t}}] \in \ZZ_p^{N_{\rm r} \times N_{\rm t}}$ as a generating matrix of a linear block code $\Cc$ of block length $N_{\rm r}$ and code rate $\frac{N_{\rm t}}{N_{\rm r}}$.
Specifically, let $\cv=[c_1,\ldots,c_{N_{\rm t}}]^{\transp}$ denote the channel inputs from the $N_{\rm t}$ transmit antennas.
The receiver knows the channel matrix $\Qm$ and can therefore create all possible codeword vectors of length $N_{\rm r}$ in a code $\Cc$ as
\begin{equation*}
\Cc=\left\{\bigoplus_{\ell=1}^{N_{\rm t}} \qv_{\ell} c_{\ell} : c_{\ell} \in \ZZ_p\right\}.
\end{equation*}
After acquiring a set of $p^{N_{\rm t}}$ codewords, the receiver can compute the minimum distance of the $\Cc$ by finding a minimum weight non-zero codeword in $\Cc$.
The minimum distance $d_{\rm min}(\Qm)$ of the codeword set is completely determined by the channel matrix $\Qm$.
Therefore, the goodness of the channel matrix can be defined with its associated minimum distance in the finite-field MIMO system.

Now, we explain the proposed scheme that uses linear block codes. The receiver observes
\begin{equation}
    \uv = \Qm\cv \oplus \tilde{\zv}.
\end{equation}
Then it can be rewritten as
\begin{equation}
    \uv = \rv \oplus \tilde{\zv} \in \ZZ_{p}^{N_{\rm r}},
\end{equation}
where the encoding is performed with the generator matrix $\Qm$, i.e.,
\begin{equation}
    \rv = \Qm \cv \in \Cc.
\end{equation}
We can define a non-linear function $f(\cdot): \ZZ_{p}^{N_{\rm r}} \rightarrow \ZZ_{p}^{N_{\rm t}}$ as the minimum distance decoding, i.e.,
\begin{equation}
    \hat{\uv} \in \ZZ_{p}^{N_{\rm t}} = f(\uv) = \argmin_{\xv \in \ZZ_{p}^{N_{\rm t}}} d_\mathsf{H}(\Qm\xv , \uv),
\end{equation}
where in repetition code, minimum distance (MD) decoding is equivalent to  majority decoding.
Because the achievable rate of the MD decoding can be fully characterized by a minimum distance of an underlying code and block length,
    we let $\mathbb{P}(\hat{\uv} \neq \cv) = P_{e}(N_{\rm r}, d_{\rm min}(\Qm))$ represent its error probability.

As in Section~\ref{subsec:diversity}, we can now create $N_{{\rm t}}$ parallel BSCs in which each sub-channel $m$ is
\begin{equation}
    \hat{u}_{m} = c_{m} \oplus \zeta_m
\end{equation}
for $m \in \{1,...,N_{{\rm t}}\}$.
By applying a capacity-achieving outer code to each sub-channel independently, an achievable sum-rate is
\begin{align}
    R_{\mbox{\tiny LBC}}&=N_{{\rm t}}\log{p}  \nonumber\\
    &-\sum_{i=1}^{N_{{\rm t}}}(\mathsf{H}_2(\mathbb{P}(\zeta_i \neq 0)) + \mathbb{P}(\zeta_i \neq 0)\log{(p-1)}).\label{eq:ach_sep}
\end{align}
This value can be computed numerically, but it is too complicated to compute the error probability of each sub-channel (i.e., $\mathbb{P}(\zeta_i \neq 0)$) as a closed-form expression.

Therefore, we consider another decoding approach for which we can derive a closed-form expression of an achievable rate as follows.
We let $\mathbb{P}(\hat{\uv} \neq \cv) = P_{e}(N_{\rm r}, d_{\rm min}(\Qm))$ denote the error-probability of the above MD decoding.
Note that $P_{e}(N_{\rm r}, d_{\rm min}(\Qm))$ is easily computable.
We yield a (scalar) point-to-point channel defined over an extension field $\FF_{p^{N_{\rm t}}}$ by using the one-to-one mapping $\Phi: \ZZ_{p}^{N_{\rm t}} \rightarrow \FF_{p^{N_{\rm t}}}$ as
\begin{equation}
    \Phi(\hat{\uv}) = \Phi(\cv) \oplus \Phi(\zetav),
\end{equation}
where  $\mathbb{P}(\Phi(\zetav) \neq 0) = P_{e}(N_{\rm r}, d_{\rm min}(\Qm))$.
Exactly following the proof technique in Theorem~\ref{thm:SIMO}, we can obtain a lower bound of the achievable rate of the above channel  as
\begin{align*}
    R&=N_{{\rm t}}\log{p} - \mathsf{H}(\Phi(\zetav))\\
    &\geq N_{\rm t}\log{p} - \mathsf{H}_2(P_{e}(N_{\rm r},d_{\rm min}(\Qm))) \\
    &\qquad\qquad~~ - P_{e}(N_{\rm r},d_{\rm min}(\Qm))\log{(p^{N_{\rm t}}-1)},
\end{align*}
which is obtained only using  $P_{e}(N_{\rm r}, d_{\rm min}(\Qm))$.
Based on this result, we have:

%%%%%%%%%%%%%%%%%%%%%%%%%%%%%%%
\begin{theorem}\label{thm:MIMO}
    Consider a MIMO channel with a full-rank channel matrix $\Qm \in \ZZ_{p}^{N_{\rm r} \times N_{\rm t}}$ where $\mathbb{P}(\tilde{z}_m \neq 0) = \epsilon_m$  for  $m \in \{1,...,N_{\rm r}\}$.
    Then the proposed linear block coding method achieves the sum-rate of
    \begin{align}
        R_{\mbox{\tiny eLBC}} &=N_{\rm t} \log{p} - \mathsf{H}_2(P_{e}(N_{\rm r},d_{\rm min}(\Qm))) \nonumber \\
        &\;\;\;\;\;\;\;\;\;\;\;\;\;\;\;\; - P_{e}(N_{\rm r},d_{\rm min}(\Qm))\log{(p^{N_{\rm t}}-1)},\label{eq:ach_com}
    \end{align}
    where
    \begin{align}
        &P_{e}(N_{\rm r},d_{\rm min}(\Qm)) = \nonumber \\
        &\;\;\;\;\;\;\; \sum_{k=\lfloor\frac{ d_{\rm min}(\Qm)-1}{2}\rfloor+1}^{N_{\rm r}}  \sum_{j=1}^{{{N_{\rm r}}\choose{k}} } \prod_{\ell \in \mathcal{S}_{k,j}}\epsilon_{\ell}  \prod_{m \in \mathcal{S}^c_{k,j}} (1-\epsilon_m).\label{eq:ub_err}
    \end{align}
    \end{theorem}
\begin{IEEEproof}
    Any linear code $\Cc$ with minimum distance $d_{\rm min}$ can correct at least $t=\lfloor \frac{d_{\rm min} - 1}{2} \rfloor$ errors.
    Therefore, for given channel matrix $\Qm$, the proposed scheme can decode $\cv$ correctly,
        provided the number of errors among $N_{\rm r}$ sub-channels is less than equal to $\lfloor \frac{d_{\rm min}(\Qm)-1}{2} \rfloor$.
    Based on this fact, the probability of erroneous symbol decoding is upper bounded by (\ref{eq:ub_err}).
    The rest of the proof exactly follows the proof of Theorem~\ref{thm:SIMO}.
\end{IEEEproof}

Intuitively, for a given $N_{\rm t}$, the achievable rate can increase as the minimum distance of  the generating matrix $\Qm$ tends to increase.
Because the repetition coding in Section~\ref{subsec:diversity} is a linear block code with minimum distance $N_{\rm r}$,
    Theorem~\ref{thm:MIMO} can be reduced to Theorem~\ref{thm:SIMO} by setting $N_{\rm t} = 1$ and $d_{\rm min}(\Qm) = N_{\rm r}$.\\

%%%%%%%%%%%%%%%%%%%%%%%%%%
%\begin{example} In this example, we consider a MIMO system with $N_{\rm t}=4$ and $N_{\rm r}=32$ and see the performance of coding approach as a function of minimum distance. From the Singleton bound  \cite{MacWilliams}, the minimum distance is upper bounded by  $d_{\rm min}(\Qm) \leq N_{\rm r} - N_{\rm t} + 1$. Namely, in this example,  $d_{\rm min}(\Qm) \leq 29$. From Fig.~\ref{capacity-comp}, we observe that when $d_{\rm min}(\Qm)$ is small, the coding approach performs worse than the antennal selection method. As notice before, if an outer code is applied to each sub-channel independently and its achievable sum-rate is given in (\ref{eq:ach_sep}), this approach can outperform the antenna selection even for a small minimum distance as in SIMO case.
%\hfill$\Diamond$
%\end{example}

%%%%%%%%%%%%%%%%%%%
{\bf Example 3:}\label{ex:Hamming} We consider the MIMO system with $N_{\rm t}=4$ and $N_{\rm r}=7$. Suppose that the effective binary channel matrix is a generating matrix of $[7,4]$ Hamming code. Then, the received observation can be written as
\begin{align*}
\uv =\underbrace{\left[%
\begin{array}{cccc}
1   & 0 & 0 & 0 \\
0 & 1 & 0 & 0 \\
 0 & 0 & 1   & 0 \\
 0 & 0 & 0   & 1 \\
 1 & 1 & 0   & 1 \\
 1 & 0 & 1   & 1\\
 1 & 1 & 1   & 0 \\
\end{array}%
\right]}_{\Qm}\cv + \tilde{\zv},
\end{align*} where $d_{\rm min}(\Qm) = 3$ \cite{MacWilliams}.
In this example, the receiver is able to correctly decode a codeword $\cv$ if the number of erroneous sub-channels is at most one.
Suppose that a transmit signal is $\cv=[0,0,0,0]^{\transp}$ and a noise vector is $\tilde{\zv}=[0,1,0,0,0,0,0]^{\transp}$.
Then the receiver observes $\uv=[0,1,0,0,0,0,0]^{\transp}$ and performs the MD decoding as
\begin{equation}
\hat{\rv} = \argmin_{\cv \in \Cc}d_\mathsf{H}(\rv, \uv).
\end{equation}
This decoding obviously finds a correct codeword $\hat{\rv} = [0,0,0,0,0,0,0]^{\transp}$ because $d_{h}(\hat{\rv},\uv) = 1$, and because $d_{\rm min}(\Qm) = 3$,
no other codeword $\rv$ has $d_\mathsf{H}(\rv,\uv) \leq 1$.
However, for noise vectors $\tilde{\zv}$ with $w_\mathsf{H}(\tilde{\zv})>1$, the MD decoding finds a wrong codeword with the Hamming weight $3$.
For example, if $\tilde{\zv}=[0,1,0,0,0,0,1]^{\transp}$, this decoding finds a wrong codeword $\hat{r} = [0,1,0,0,1,0,1]^{\transp}$.\\

{\bf Remark 2:} The proposed approach is useful when $N_{\rm t}$ and $N_{\rm r}$ are very large and $p\geq 3$, in which the computational complexity of the conventional ML detection methods \cite{Choi2016} grows exponentially with both $N_{\rm t}$ and $p$, whereas the computational complexity of the proposed detection methods increases linearly with both $N_{\rm t}$ and $p$,
    because they estimate a symbol vector over the equivalent linear MIMO channel, instead of over the non-linear MIMO channel.

%%%%%%%%%%%%%%
\section{Numerical Results}

For the simulation, we consider a random effective channel matrix $\Qm \in \ZZ_{p}^{N_{{\rm r}} \times N_{{\rm t}}}$,
    where each element of $\Qm$ can take a value from $\{0,...,p-1\}$ uniformly and independently from other elements.
For each sub-channel $m$, we assume that
\begin{align*}
&\mathbb{P}(\tilde{z}_m = 0) = 1 - \epsilon_m\\
&\mathbb{P}(\tilde{z}_m= k) = \frac{\epsilon_m}{p-1} \mbox{ for } k \in \{1,...,p-1\}.
\end{align*}
Thus, each sub-channel is specified by one parameter $\epsilon_m$.
We consider an average achievable sum rate, where the average is with respect to a random effective channel matrix $\Qm$.
For the decoding methods, we consider the antenna selection with successive coding (AS) in Section~\ref{subsec:AS}, and two linear block coding methods in Section~\ref{subsec:LBC}
    where one (LBC) uses $N_{{\rm t}}$ outer codes and the other (eLBC) uses one outer code over an extension field.
The achievable sum rates of AS, LBC, and eLBC are given in (\ref{eq:ach_AS}),  (\ref{eq:ach_sep}), and  (\ref{eq:ach_com}), respectively.  \\

%%%%%%%%%%%%%%%%%%%%%%%%%%
\begin{figure}
\centerline{\includegraphics[width=9cm]{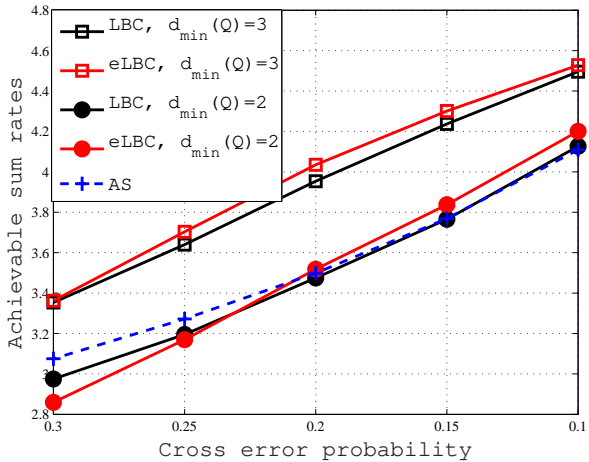}}
\caption{$N_{\rm t} = 2$, $N_{\rm r}=5$ and $p=5$. Achievable sum rates of the proposed methods as a function of  $d_{\rm min}(\Qm)$. The achievable sum rate of AS is independent from $d_{\rm min}(\Qm)$.}
\label{rate-md}
\end{figure}
%%%%%%%%%%%%%%%%%%%%%%%%%%%

%%%%%%%%%%%%%
{\bf Example 4:} {\em (Effect of minimum distance)} In this example, we see the impact of minimum distance of $\Qm$ on achievable sum rates.
Consider the MIMO system with $N_{{\rm t}}=2$ and $N_{{\rm r}}=5$.
Let $\Hc$ denote the sample space containing all possible realizations of $\Qm$ and $\Ac_{\ell} = \{\Qm \in \Hc: d_{\rm min}(\Qm) = \ell\} \subseteq \Hc$.
Here, we compute the conditional achievable sum rates for which the average is with respect to a random matrix $\Qm$ with $d_{\rm min}(\Qm)=\ell$.
The corresponding numerical results are provided in Fig.~\ref{rate-md}.
It is observed that both coding methods can yield a better achievable sum rate than AS and the performance gap increases as $d_{\rm min}(\Qm)$ increases.\\

%%%%%%%%%%%%%%%%%%%%%%%%%%%%%
\begin{figure}
\centerline{\includegraphics[width=9cm]{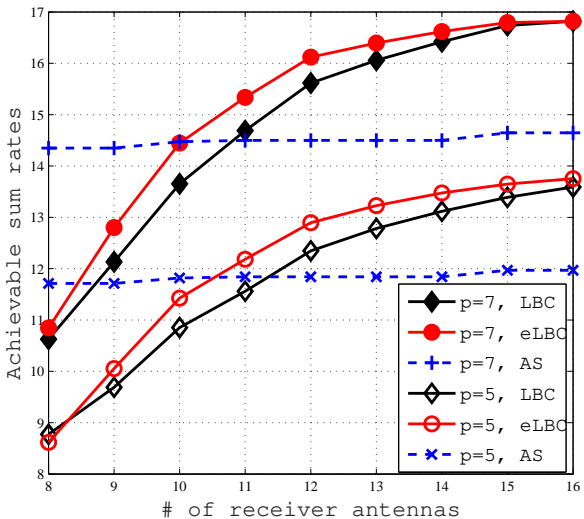}}
\caption{$N_{\rm t}=6$. Achievable sum rates of the proposed methods as a function of number of receiver antennas.}
\label{nt_6}
\end{figure}

%%%%%%%%%%%%%
{\bf Example 5:} {\em (Effect of number of receiver antennas)}
We consider the MIMO system with $N_{{\rm t}} =6$ and $N_{{\rm r}} \geq 16$ where  $\epsilon_{i}$ is a random variable which can take a value between $0.05$ and $0.15$ uniformly and independently of other cross-probabilities.
The corresponding numerical results are provided in Fig.~\ref{nt_6}.
We first observe that both LBC and eLBC can give higher achievable sum rates than AS when $N_{\rm r}$ is large enough.
This result occurs because when $N_{\rm r}$ is small the minimum distance of $\Qm$ is small (in this example, $d_{\rm min}(\Qm) < 3$ for $N_{\rm r} < 12$),
so the achievable rate of the coding method is inferior.
Thus, we much consider $N_{\rm r}$ and $N_{\rm t}$ when choosing a MIMO detection method.\\

From Example 4 and 5, we can see that the achievable sum rate of LBC less than 0.5 bits lower than that of eLBC;
therefore we recommend use of LBC in practice because it has much lower complexity than eLBC.

%%%%%%%%%%%%%%%%
\section{Conclusion}
In this paper, we have presented a novel detection approach for the MIMO system with low-resolution ADCs by exploiting coding theory.
In particular, by using the proposed quantizer and lattice modulation-demodulation technique,
    we have created an equivalent linear MIMO system with finite-field input-output values.
Then, applying algebraic coding theory, we have introduced a set of detection methods that apply to different antenna configurations, and have characterized the corresponding achievable rates.

One possible future work would be to develop channel estimation techniques that are suitable for the proposed approach.
For instance, it may be beneficial to directly estimate the effective integer channel matrix ${\bf Q}$ instead of ${\bf H}$ using pilot signals.
Although we have concentrated on single-user MIMO systems, the approach propounded in this paper can be extended to the multi-user scenario in which multiple transmitters equipped with a single antenna send independent data streams (codewords) to a receiver equipped with multiple antennas when adopting limited ADCs.
Other extensions can also be explored; for instance, in the SIMO case, one can further improve the achievable rates by using a modified majority decoding rule that exploits the fact that each received signal contains different reliability information.

%%%%%%%%%%%%%%%%%%%%%%%%%%%%%%%%%%%%%%%%%%%
%%%%%%%%%%%%%%%%%%%%%%%%%%%%%%%%%%%%%%%%%%%
%%%%%%%%%%%%%%%%%%%%%%%%%%%%%%%%%%%%%%%%%%%

%%%%%%%%%%%%%%%%%%%%%%%

%%%%%%%%%%%%%%%%%%%%%%%%%%%%%%%%%%%%%%%%%%%%
\end{document}